\documentclass[]{article}
\usepackage{amsmath}
\usepackage{amsfonts}
\usepackage{amssymb}
\usepackage{graphicx}
\usepackage{amsfonts}
\usepackage[english]{babel}
\usepackage[utf8]{inputenc}
\usepackage{times}
\usepackage[T1]{fontenc}
\usepackage{multirow}
\usepackage{xcolor}
\usepackage{hyperref}
\usepackage{cite}

\begin{document}

\title{Barbero-Immirzi Value from Experiment}

\author{Leonid Perlov, \\
Department of Physics, University of Massachusetts, Boston, USA\\
leonid.perlov@umb.edu
}

\maketitle

\begin{abstract}
We consider General Relativity as a limit case of the Scalar-Tensor theory with Barbero-Immirzi field when the field tends to a constant. We use Shapiro time delay experimental value of $1/w(x) = (2.1 \pm 2.3)10^{-5}$ provided by the Cassini spacecraft to find the present Barbero-Immirzi parameter value.
\end{abstract}
This is a short note on obtaining Barbero-Immirzi value from the experimental data by using the Parametric Post Newtonian framework  \cite{Will}. Currently there are only two types of theories that still agree with experiment: General Relativity (GR) and the Scalar-Tensor theory \cite{Brans, Brans2}. The Barbero-Immirzi  parameter plays a major role in Loop Quantum Gravity and in Ashtekar's GR formulation \cite{Ashtekar, RovelliBook, Thiemann}. This paper determines Barbero-Immirzi parameter's value $\beta$ directly from experimental data. The previous attempts to restrain the Barbero-Immirzi parameter can be found in \cite{Perez} - \cite{Shaposhnikov1}. Barbero-Immirzi parameter appears in the Holst action \cite{Holst} that is used in Loop Quantum Gravity and in Ashtekar's GR formulation as a coefficient of the topological term that vanishes due to the first Bianchi identity:
\begin{equation}
\label{1}
S = \frac{1}{16\pi G} \int d^4x \; e e^{\mu}_I e^{\nu}_J (R^{IJ}_{\;\;\mu\nu} - \frac{\beta}{2}\epsilon^{IJ}_{\;\;\;KL}R^{KL}_{\;\;\mu\nu}) . 
\end{equation}
Thus, it is impossible to obtain its value directly from the classical GR theory. Formally Barbero-Immirzi parameter can be real or complex valued. In this paper we consider GR as a limit of the Scalar-Tensor theory with Barbero-Immirzi field (BI) and non-zero torsion. When torsion tensor tends to zero, BI field tends to a constant, as it follows from \cite{Mercuri}, also see details in Appendix B. The Barbero-Immirzi field theory was introduced and studied in \cite{Mercuri, Krasnov, Taveras, Montani}, and \cite{Oscar}. Its action obtained from ($\ref{1}$) when $\beta(x)$ is a field is as follows \cite{Mercuri}:
\begin{equation}
\label{12}
S = \frac{1}{16\pi G}  \int \sqrt{-g}\;  d^4x \; \left[R - \frac{3}{2} \left(\frac{1}{1+ \beta^2(x)}\right) \;  \partial_{\mu} \beta(x) \partial^{\mu} \beta(x) \right] , 
\end{equation}
where $R$ is a torsion-free Riemann scalar. \\[2ex]
The Scalar-Tensor theories are rare types of theories besides GR that are still in agreement with the experimental data \cite{Will, Will2}.
The Scalar-Tensor theory action written in ''Einstein frame'' contains a scalar field $\phi(x)$ and a coupling function $w(\phi(x))$ \cite{Will, Brans, Weinberg}:
\begin{equation}
\label{22}
S = \frac{1}{16\pi G} \int \sqrt{-\tilde{g}} \; d^4x \; \left[ \tilde{R}  - \frac{3+ 2w(\phi(x))}{2{\phi^2(x)}}  \partial_{\mu} \phi(x)\partial^{\mu} \phi(x) \right] , 
\end{equation}
where $\tilde{R}$ is a Riemann curvature after the conformal transformation $g_{\mu\nu} =  \tilde{g}_{\mu\nu}/\phi(x)$.\\[2ex]
In the limit, when $w(\phi(x))$ is constant and tends to infinity, the theory becomes pure GR \cite{Brans, Weinberg}: $\phi(x) = const + O(\frac{1}{w})$,  $R_{\mu\nu} - \frac{1}{2}g_{\mu\nu}R = -8\pi G T_{\mu\nu} + O(\frac{1}{w})$\\[2ex]
Moreover, one can use the scaling $g_{\mu\nu} =  \tilde{g}_{\mu\nu}/{\phi^2(x})$ instead of  $g_{\mu\nu} =  \tilde{g}_{\mu\nu}/\phi(x)$ (see Appendix A for details) to present it in the form:
\begin{equation}
\label{577}
S = \frac{1}{16\pi G} \int \sqrt{-\tilde{g}} \; d^4x \; \left[ \frac{\tilde{R}}{\phi(x)}  - \frac{6+w(\phi(x)}{{\phi^3(x)}}  \partial_{\mu} \phi(x)\partial^{\mu} \phi(x) \right] ,
\end{equation}
We then perform a conformal transformation  $g_{\mu\nu} =  \tilde{g}_{\mu\nu}/{\beta(x)}$  in ($\ref{12}$) to obtain:
\begin{equation}
\label{323}
S = \frac{1}{16\pi G}  \int \sqrt{-\tilde{g}}\;  d^4x \; \left[\frac{\tilde{R}}{\beta(x)} - \frac{3}{2} \left(\frac{1}{1+ \beta^2(x)}+ \frac{1}{\beta^3(x)}\right) \;  \partial_{\mu} \beta(x) \partial^{\mu} \beta(x) \right] .
\end{equation}	
(\ref{577}) and (\ref{323}) are equal when the second terms are equal and the scales are equal: $\phi^2(x) = \beta(x)$, since in one case we used $\phi^2(x)$ scaling, while in the other $\beta(x)$. Thus, we have obtained a system of two equations:
\begin{multline}
\label{324}
\begin{aligned}
 \phi^2(x) = \beta(x), \\[2ex]
\frac{3}{2}\left(\frac{1}{(1+{\beta^2(x)})} + \frac{1}{\beta^3(x)}\right) = \frac{6 + w(\beta(x))}{{\phi^3(x)}} .
\end{aligned}
\end{multline}
After substituting the first equation into the second and simplifying, we obtain the following equation for $\phi(x)$:
\begin{equation}
\label{325}
(12 + 2w(\phi(x))\;\phi^7(x) - 3\phi^6(x) - 3\phi^4(x) + (12+2w(\phi(x))\;\phi^3(x) - 3 = 0.
\end{equation}	
When $w(x)$ tends to infinity, we receive in the limit equation by dividing by $12 + 2w(x)$
\begin{equation}
\lim_{w \to \infty} \left[ \phi^7(x) - \frac{3\phi^6(x)}{(12 + 2w(\phi(x))} - 3\frac{\phi^4(x)}{(12 + 2w(\phi(x))}+ \phi^3(x) - \frac{3}{(12 + 2w(\phi(x))} = 0\right].
\end{equation}
\begin{equation}
\label{333}
\phi^7(x) + \phi^3(x) = 0.
\end{equation}
or
\begin{equation}
\label{334}
\phi^3(x) (\phi^4(x) +1) = 0.
\end{equation}\\
Recalling from (\ref{324}) that $\phi^2(x) = \beta(x)$, we rewrite ($\ref{334}$) with $\beta(x)$
\begin{equation}
\label{335}
\beta^{3/2}(x) (\beta^2(x) +1) = 0.
\end{equation}
The asymptotic solutions are:
\begin{equation}
\label{8}
\beta(x) = \pm i \;\;\; \mbox{or}  \;\;\; \beta(x) = 0
\end{equation}
By solving (\ref{325}) numerically with $w(\phi(x)) =  10^5/(2.1\pm 2.3)$, we obtain the value, which is very close to the asymptotic one:
\begin{equation}
\beta(x) = \pm (i + 1.27 \;10^{-5}) \pm (0.1i + 0.06)10^{-5}
\end{equation}
As it seen from ($\ref{334}$), each of two roots of ($\ref{325}$) is of a multiplicity two, while the zero root is of a multiplicity three. Of course, we disregard the zero root, since we divide by $\phi(x)$ and $\beta(x)$ in  our original equation ($\ref{324}$)\\[2ex]
Let us now discuss the experimental data obtained for $w(\phi(x))$. The most recent experimental value of $w(\phi(x))$ is $ 10^5/(2.1\pm 2.3)$. It was provided by Shapiro time delay data in 2003 from the Cassini spacecraft on its way to Saturn \cite{Bertotti, Freire, Will}. It's worth mentioning that back in 1972 this value was much lower, equal to 6 \cite{Brans, Will2}, reaching thousands at the end of 70th, and order of $10^5$ today. The ongoing BepiColombo mission to Mercury, launched in 2018, will further increase Cassini's result 4 folds in July 2022 when it will experience the first solar conjunction \cite{Imperi}. \\[2ex] 
For the combined graph of the different experimental data we refer to  \cite{Freire}. The values provided by the most recent binary pulsars are not competitive with the solar-system Shapiro time delay measurement provided by Cassini due to the near equality of the star masses suppressing dipole radiation \cite{Will}.\\[2ex]
The bigger $w(\beta(x))$ the closer is the Scalar-Tensor theory to GR. When $w(\beta(x))$ is constant, which is the original Brans-Dicke theory \cite{Brans}, the limit $w \rightarrow \infty$ implies $\beta(x) =  const + O(\frac{1}{w})$ \cite{Weinberg}.  As we see from ($\ref{8}$) the limit of $\beta = \beta(x) \rightarrow const $ in this case is $\pm i$, which corresponds to the Ashtekar sefl-dual GR formalism. At the end we would like to address the solution uniqueness. Indeed, by looking at  ($\ref{12}$) we see that when $\beta(x)$ is any constant, not necessarily $\pm i$, the action becomes GR. However,  if we want the theory to be in agreement with the experiment all the time, while going to a limit, then the limit value of $\beta(x)$ cannot be a random constant, but necessary $\pm i$. To conclude, today's value of the Barbero-Immirzi parameter detected from experiment is $ \beta(x) = \pm (i + 1.27 \;10^{-5}) \pm (0.1i + 0.06)10^{-5}$, with the expected BepiColombo improvement it will become $\beta(x) = \pm (i + 2.95\;10^{-6}) \pm (0.0.5i + 0.16)10^{-6}$, and with further experiments  it will be going closer and closer to Ashtekar's limit value $\pm i$. \\[2ex]
\textbf{Appendix A - Scalar-Tensor Action Scaling}\\[2ex]
 The original form of the Scalar-Tensor theory action is as follows  \cite{Brans, Will3, Weinberg}:
 \begin{equation}
\label{590}
S = \frac{1}{16\pi G} \int \sqrt{-{g}} \; d^4x \; \left[ \phi(x) R  - \frac{w(\phi(x))}{{\phi(x)}} g^{\alpha\beta} \partial_{\alpha} \phi(x)\partial_{\beta} \phi(x) \right] ,
\end{equation}
Let us make the transform used in this paper: $g_{\alpha\beta} = \tilde{g}_{\alpha\beta}/\phi^2(x)$. We have correspondingly $g^{\alpha\beta} = \phi^2(x)\tilde{g}^{\alpha\beta}$, the volume will transforms as $\sqrt{-g} = \sqrt{-\tilde{g}}/\phi^4(x)$, and the Riemann scalar as \cite{Will3}:
\begin{equation}
R = \phi^2(x) \left(\tilde{R} + 6\frac{\partial_{\alpha}\phi(x)}{\phi^2(x)} - 6\frac{\tilde{g}^{\alpha\beta}\partial_{\alpha}\phi(x)\partial_{\beta}\phi(x)}{\phi^2(x)}\right)
\end{equation}
By substituting all these components into ($\ref{590}$) and eliminating $6\frac{\partial_{\alpha}\phi(x)}{\phi^2(x)}$ by the Gauss theorem converting it to a surface integral \cite[p.~740]{Will3}  we obtain:
\begin{equation}
\label{589}
S = \frac{1}{16\pi G} \int \sqrt{-\tilde{g}} \; d^4x \; \left[ \frac{\tilde{R}}{\phi(x)}  - \frac{6+ w(\phi(x))}{{\phi^3(x)}}  \tilde{g}^{\alpha\beta}\partial_{\alpha} \phi(x)\partial_{\beta} \phi(x) \right] ,
\end{equation}\\[2ex]
\textbf{Appendix B - BI Field is Constant when the Torsion Tensor is Zero}\\[2ex]
We repeat here the equations 8, 15-17 of \cite{Mercuri} to demonstrate this point. 
The torsion tensor can be split into irreducible components:
\begin{equation}
\label{B1}
T_{\mu\nu\rho} = \frac{1}{3}(T_{\nu}g_{\mu\rho} - T_{\rho}g_{\mu\nu}) - \frac{1}{6}\epsilon_{\mu\nu\rho\sigma}S^{\sigma} + q_{\mu\nu\rho},
\end{equation}
where $T_{\mu} = T^{\nu}_{\mu\nu}$, $S_{\mu} = \epsilon_{\nu\rho\sigma\mu}T^{\nu\rho\sigma}$, $q_{\mu\nu\rho}$ - the antisymmetric tensor, such that $q^{\nu}_{\rho\nu} = 0= \epsilon^{\mu\nu\rho\sigma}q_{\mu\nu\rho}$\\[2ex]
By varying the Holst action with Barbero Immirzi being a field, and using the fore-mentioned components of the torsion tensor, we obtain the system of equations:
\begin{equation}
\label{B2}
S_{\mu} = \frac{6}{1 + \beta^2(x)}\partial_{\mu}\beta \;,
\end{equation}
\begin{equation}
\label{B3}
T_{\nu} = \frac{3}{2} \frac{\beta(x)}{1 + \beta^2(x)}\partial_{\nu}\beta \;,
\end{equation}
\begin{equation}
\label{B4}
q_{\mu\nu\rho} = 0 \;.
\end{equation}
It follows then from $(\ref{B1}), (\ref{B2}), (\ref{B3})$, and $(\ref{B4})$ that the BI field is constant if and only if the torsion tensor is zero. 
\\[4ex]
\textbf{Acknowledgments}\\[2ex]
Dedicated to the memory of my beloved father who was very supportive and much interested in the fate of this paper.\\[2ex]
I am very grateful to Michael Bukatin for reviewing this note and for his supportive enthusiastic spirit. \\[2ex]

\end{document}